\documentclass[amsmath, amssymb, aip, apl, reprint, twocolumns, superscriptaddress]{revtex4}
\usepackage{graphicx}
\usepackage{dcolumn}
\usepackage{physics}
\usepackage{braket}
\usepackage{bm}
\usepackage{color}
\usepackage{hyperref}
\hypersetup{colorlinks=true,citecolor=blue,linkcolor=blue, urlcolor=blue}
\newcommand{\kk}{\mathbf{k}}
\newcommand{\QQ}{\mathbf{Q}}
\renewcommand{\qq}{\mathbf{q}}

\renewcommand{\vec}[1]{\mathbf{#1}}
\begin{document}

\title{Precise Radiative Lifetimes in Bulk Crystals from First Principles:\\ The Case of Wurtzite GaN}
%
%
\author{Vatsal A. Jhalani}
\affiliation{Department of Applied Physics and Materials Science, California Institute of Technology, 1200 E. California Blvd., Pasadena, California 91125, USA.}
\author{Hsiao-Yi Chen}
\affiliation{Department of Applied Physics and Materials Science, California Institute of Technology, 1200 E. California Blvd., Pasadena, California 91125, USA.}
\affiliation{Department of Physics, California Institute of Technology, 1200 E. California Blvd., Pasadena, \mbox{California 91125, USA.}}
\author{Maurizia Palummo}
\affiliation{Dipartimento\! di\! Fisica, \!\!Universit\`a di Roma \!\!``Tor Vergata'' and INFN, \mbox{00133 Roma, Italy}} 
\author{Marco Bernardi}
\email[Please E-mail correspondence to: ]{bmarco@caltech.edu}
\affiliation{Department of Applied Physics and Materials Science, California Institute of Technology, 1200 E. California Blvd., Pasadena, California 91125, USA.}



\begin{abstract}
Gallium nitride (GaN) is a key semiconductor for solid-state lighting, but its radiative processes are not fully understood. 
Here we show a first-principles approach to accurately compute the radiative lifetimes in bulk uniaxial crystals, focusing on wurtzite GaN. 
Our computed radiative lifetimes are in very good agreement with experiment up to 100 K. 
We show that taking into account excitons (through the Bethe-Salpeter equation) and spin-orbit coupling to include the exciton fine structure is essential for computing accurate radiative lifetimes. 
A model for exciton dissociation into free carriers allows us to compute the radiative lifetimes up to room temperature. 
Our work enables precise radiative lifetime calculations in III-nitrides and other anisotropic solid-state emitters. 
\end{abstract}

\pacs{}

\maketitle

%
%
Semiconductor light-emitting diodes (LEDs) are the preferred light source for many applications. In LEDs, electroluminescence converts electron and hole carriers into emitted photons. 
This radiative recombination process depends on material properties such as the band structure, dielectric screening, and optical excitations. 
In a simplified picture, free electrons and holes in band states recombine to emit light. Yet, in many solid-state emitters electron-hole interactions are strong enough to form excitons 
(bound electron-hole states)~\cite{knox1963}, and the main process leading to light emission is exciton radiative recombination.  
Gallium nitride (GaN) is widely employed for efficient light emission~\cite{LEDprospects}, and it has been investigated extensively in crystalline, thin film, and heterostructure forms, both to understand its physical properties and 
to improve LED devices. Even though the exciton binding energy is rather weak  in GaN (of order 20 meV~\cite{MuthGaNBE}), accurately computing its absorption spectrum requires taking into account excitonic effects~\cite{Laskowski2005}, 
so one expects that excitons also play a role in light emission.\\
\indent
The radiative properties of GaN have remained the subject of debate~\cite{Harris1995, Chen1996, MuthGaNBE, Im1997, Brandt1998, Vatsal2017}. 
Investigations of radiative processes require photoluminescence (PL) spectroscopies or device experiments on pure samples. Since GaN films are typically grown epitaxially, and their doping is nontrivial, 
these measurements are affected by sample purity and competing non-radiative processes due to defects and interfaces~\cite{WolfeExcTherm}. 
In addition, typical theoretical treatments of radiative lifetimes employ simplified empirical methods that can only qualitatively interpret, or just fit, experimental data~\cite{Ridley}. 
Accurate first-principles calculations of the \textit{intrinsic} radiative properties of a GaN crystal would be highly desirable 
as they would serve as a benchmark for interpreting PL measurements and for guiding microscopic understanding and device design. 
Isolated examples of \textit{ab initio} radiative lifetime calculations in bulk materials exist~\cite{zhang2018perov,kioupakis2013nitride}, but they neglect key factors such as excitonic effects, the material anisotropy, and temperature dependence dictated by dimensionality.\\
\indent
We have recently shown a general first-principles approach that includes these physical features and can be applied broadly to systems ranging from isolated emitters to bulk crystals~\cite{Chen0D3D}. 
Its application to carbon nanotubes \cite{Spataru2005}, two-dimensional materials \cite{PalummoTMDC, ChenTMDC, GaoTMDC}, and recently gas phase molecules has shown results in very good agreement with experiments. 
The method is based on the solutions of the \textit{ab initio} Bethe-Salpeter equation (BSE) \cite{Strinati1984,RohlfingBSE}, which can correctly treat excitons but is computationally expensive, 
plus Fermi's golden rule to obtain the radiative lifetimes and their temperature dependence. 
For comparison, the widely used independent-particle picture (IPP) of light emission, which does not take excitons into account, is straightforward to compute, but as we show below it makes large errors on the radiative lifetimes. 
For bulk crystals, we are not aware of \textit{ab initio} radiative lifetime calculations that properly include excitons other than our recent work, which focused on isotropic crystals~\cite{Chen0D3D}. 
However, GaN is isotropic only in the hexagonal basal plane, while the properties along the $c$-axis are different. 
Its anisotropic optical properties cannot be taken into account in current \textit{ab initio} radiative lifetime calculations.\\ 
\indent
%
%
Here, we derive a first-principles approach to compute the radiative lifetimes of a uniaxial bulk crystal, and apply it to wurtzite GaN. 
The computed radiative lifetimes are in very good agreement (within a factor of two) with experiment up to 100~K,  
and we include thermal exciton dissociation to retain quantitative accuracy up to room temperature. 
In spite of the weak exciton binding energy in GaN, we show that including excitons is essential for quantitative 
accuracy as it improves substantially the agreement with experiment compared to IPP calculations. 
We also show that including spin-orbit coupling (SOC) and the related exciton fine structure is important in spite of the weak SOC in GaN. %
Our work advances the study of light emission in III-nitrides and anisotropic light emitters.\\
\indent
%
%
Our discussion in Ref.~\onlinecite{Chen0D3D} forms the basis for deriving the radiative lifetimes of a uniaxial crystal. 
The dielectric tensor of a uniaxial bulk crystal, 
\begin{equation} \label{eq:diel}
\epsilon_r = \text{diag}(\epsilon_{xy},\epsilon_{xy},\epsilon_{z}),
\end{equation}
is isotropic in the basal hexagonal plane, and different along the principal crystal axis (the $z$ direction). 
For a given photon wavevector $\qq$, there are two non-degenerate propagating modes as solutions to Maxwell's equations, each corresponding to one of the two photon polarizations \cite{Glauber1991}.  
We call the first solution the ``in-plane'' (IP) mode since its polarization vector sits in the $xy$-plane, and the second solution the ``out-of-plane'' (OOP) mode, 
which sees the anisotropy of the material and as shown below has a more complicated expression than in the isotropic case~\cite{Chen0D3D}.
The polarization vectors $\mathbf{e}$ and frequencies $\omega$ of the two modes are obtained by solving the equation of motion of the vector potential in the dielectric material \cite{Chen0D3D}. 
For the uniaxial case, we get: 
\begin{widetext}
\vspace{-15pt}
\begin{align} 
\label{eq:IP}
\frac{\omega_\text{IP}}{c} &= \sqrt{\frac{q^2}{\epsilon_{xy}}}, \,\,\,\,\,\,\,\,\,\,\,\,\,\,\,\,\,\,\,\,\,\,\,\,\,\,\,\,\,\,\,\,\,\,\,   \mathbf{e}_\text{IP} = \frac{1}{\sqrt{\epsilon_{xy}}}\left( \frac{q_y}{q_{xy}},-\frac{q_y}{q_{xy}},0\right) \\
\label{eq:OOP}
\frac{\omega_\text{OOP}}{c}&=\sqrt{\frac{\epsilon_{xy}q_{xy}^2+\epsilon_zq_z^2}{\epsilon_{xy}\epsilon_{z}}},\,\,\,\,
\mathbf{e}_\text{OOP} = \left( \frac{q_x}{q_{xy}}\sqrt{\frac{1/\epsilon_{xy}}{\left( 1 + \frac{\epsilon_{xy}q_{xy}^2}{\epsilon_zq_z^2}\right)}}, 
\frac{q_y}{q_{xy}}\sqrt{\frac{1/\epsilon_{xy}}{\left( 1 + \frac{\epsilon_{xy}q_{xy}^2}{\epsilon_zq_z^2}\right)}},
-\sqrt{\frac{1/\epsilon_{z}}{\left( 1 + \frac{\epsilon_{z}q_{z}^2}{\epsilon_{xy}q_{xy}^2}\right)}}
\right),
\end{align}
\end{widetext}
where $c$ is the speed of light and $q_{xy}^2 = q_x^2 + q_y^2$. \\
\indent
The radiative recombination rate at zero temperature for an exciton in state $S$ with center-of-mass momentum $\QQ$ can be written using Fermi's golden rule as~\cite{Chen0D3D}
\begin{equation} \label{eq:FGR}
\gamma_S(\QQ)\!=\!\frac{\pi e^2}{\epsilon_0 m^2 V} \sum_{\lambda\qq} \frac{1}{\omega_{\lambda\qq}} \left | \vec{e}_{\lambda\qq} \cdot \vec{p}_S(\QQ) \right |^2 \delta (E_S(\QQ) - \hbar \omega_{\lambda\qq}),
\end{equation}
where V is the volume of the system and $\epsilon_0$ the vacuum permittivity, and the sum over $\lambda$ adds together the contributions from the IP and OOP modes.  
The transition dipoles of the exciton, $\vec{p}_S(\vec{Q})$, are computed using the velocity operator to correctly include the nonlocal part of the Hamiltonian \cite{Chen0D3D, Sangalli2017}.
Since the values of $\QQ$ relevant for light emission are small, we approximate the transition dipoles
as $\vec{p}_S(\QQ) \!\approx\! \vec{p}_S(0)$, and obtain them, together with the exciton energies $E_S(0)$, by solving the BSE at $\QQ=0$. 
We substitute into Eq.~(\ref{eq:FGR}) the two solutions in Eqs.~(\ref{eq:IP}) and (\ref{eq:OOP}), use momentum conservation, which fixes the emitted photon wave vector to $\qq = \QQ$, 
and obtain the radiative recombination rate at zero temperature for each exciton state $S$ in a uniaxial bulk material (we put $Q_{xy}^2 = Q_x^2 + Q_y^2$): 
\begin{widetext}
\vspace{-15pt}
\begin{align}
\begin{split} 
\label{gamma3D}
\gamma_S(\QQ) = 
\frac{\pi e^2}{\epsilon_0 m^2 V}
&\left [ \frac{\sqrt{\epsilon_{xy}}}{cQ} \left | \frac{1}{\sqrt{\epsilon_{xy}}} \frac{p_{Sx}Q_y-p_{Sy}Q_x}{Q_{xy}} \right |^2_\text{IP} \! \delta \!\left(E_S(Q)-\frac{\hbar cQ}{\sqrt{\epsilon_{xy}}}\right) 
\,\,+\,\, \frac{\sqrt{\epsilon_{xy}\epsilon_z}}{c\sqrt{\epsilon_{xy}Q^2_{xy}+\epsilon_zQ^2_z}} \,\times \right . \\
&\,\,\,\, \left . \left | \frac{Q_xp_{Sx}+Q_yp_{Sy}}{Q_{xy}} \sqrt{\frac{1/\epsilon_{xy}}{1+\frac{\epsilon_{xy}Q_{xy}^2}{\epsilon_zQ_z^2}}} - p_{Sz} \sqrt{\frac{1/\epsilon_z}{1+\frac{\epsilon_zQ_z^2}{\epsilon_{xy}Q_{xy}^2}}} \right |^2_\text{OOP}
\!\! \delta \! \left ( E_S(Q) - \frac{\hbar c \sqrt{\epsilon_{xy}Q^2_{xy} + \epsilon_zQ_z^2}}{\sqrt{\epsilon_{xy}\epsilon_z}} \right)
\right ].
\end{split}
\end{align}
\end{widetext}
Assuming that the exciton momentum $\QQ$ has a thermal distribution, the radiative rate of an exciton $S$ at temperature $T$ is written as the thermal average
\begin{equation} \label{eq:thermav}
\left<\gamma_S\right>(T)=\frac{\int d\QQ e^{-E_S(Q)/k_BT}\gamma_S(\QQ)}{\int d\QQ e^{-E_S(Q)/k_BT}},
\end{equation}
where $k_B$ is the Boltzmann constant. 
We employ an effective mass approximation for the exciton dispersion, with IP and OOP effective masses $M_{xy}$ and $M_z$, respectively, 
obtained as the sum of the electron and hole effective masses. 
Since we find from the BSE that the lowest exciton states are composed of transitions from the two heavy-hole bands, we approximate the hole mass as the average of the two heavy-hole masses.
The exciton radiative rate at temperature $T$ is then obtained from the integral in Eq.~(\ref{eq:thermav}): 
\begin{equation}
\label{eq:gammatherm}
\left< \gamma_S \right>(T) =  \left( \frac{E_S(0)^2}{2M_{xy}^\frac{2}{3}M_z^\frac{1}{3}c^2k_BT}\right)^{3/2} \times 
\frac{\sqrt{\pi\epsilon_{xy}}e^2\hbar\left[\left(\frac{2\epsilon_z}{3\epsilon_{xy}} + 2 \right)\left( p_{Sx}^2 + p_{Sy}^2 \right) + \frac{8}{3} p_{Sz}^2 \right]}{\epsilon_0 m^2VE_S(0)^2}\,,
\end{equation}
%
where the exciton energies and transition dipoles are obtained by solving the BSE. 
The radiative lifetime is defined as the inverse radiative rate, $\left<\tau_S\right> \!=\! \left<\gamma_S\right>^{-1}$. 
Note also that Eq.~(\ref{eq:gammatherm}) reduces to the bulk isotropic case in Ref.~\onlinecite{Chen0D3D} if one puts $\epsilon_z = \epsilon_{xy}$ and $M_{xy} = M_z$.\\
\indent 
Finally, we take into account the fact that multiple exciton states can be occupied (including dark states with small transition dipoles, as is the case in GaN), 
and compute the radiative rate assuming a thermal equilibrium distribution:
\begin{equation} \label{eq:gammaeff}
\left<\gamma(T)\right> = \frac{\sum_S\left<\gamma_S\right>e^{-E_S(0)/k_BT}}{\sum_Se^{-E_S(0)/k_BT}}.
\end{equation} 
We use this thermal average, computed with the exciton radiative rates $\left<\gamma_S\right>$ in Eq.~(\ref{eq:gammatherm}), to obtain the \textit{intrinsic} radiative lifetime $\left<\gamma(T)\right>^{-1}$ in bulk wurtzite GaN.\\
\indent
%
%
We carry out first-principles calculations on a wurtzite GaN unit cell with relaxed lattice parameters. 
The ground state properties and electronic wave functions are computed using density functional theory (DFT) within the generalized gradient approximation \cite{GGA,PBEsol} with the \textsc{Quantum ESPRESSO} code \cite{QE}. 
Fully-relativistic norm-conserving pseudopotentials~\cite{ONCVPSP} generated with Pseudo Dojo~\cite{Dojo} are employed, 
in which the shells treated as valence are the $3s$, $3p$, $3d$, $4s$, and $4p$ for Ga and the $2s$ and $2p$ for N. 
A non-linear core correction \cite{NLCC} is included for all remaining core shells for both atoms. 
We compute the quasiparticle band structure in GaN~\cite{RubioGaN} with a ``one-shot'' $GW$ calculation~\cite{GWreview} with the Yambo code~\cite{Yambo1, Yambo2} using a plasmon-pole model for the dielectric function, a 25 Ry cutoff for the dielectric matrix, 300 empty bands, and a $14 \times 14 \times 10$ 
$\kk$-point grid. 
For the $GW$ band structure, we start from DFT within the local-density approximation \cite{LDA} and employ scalar-relativistic norm-conserving pseudopotentials for both Ga and N, where the $4s$ and $4p$ shells are treated as valence for Ga, and the $2s$ and $2p$ for N. A non-linear core correction is included to account for the $3d$ core states in Ga.
The BSE is solved on a $24 \times 24\times 18$ $\kk$-point grid using a 6 Ry cutoff for the static dielectric screening and the 6 highest valence bands and 4 lowest conduction bands. 
These settings are sufficient to converge the energies, transition dipoles and radiative lifetimes of the low-energy excitons, as we have verified.  
The IPP transition dipoles and energies are computed by neglecting the electron-hole interactions in the BSE.  
%
The exciton binding energy is converged by computing it with several $\kk$-point grids from $12\times 12\times 9$ to $24\ \times 24\times 18$ and extrapolating it to a vanishingly small $\kk$-point distance (\textit{i.e.}, to an infinitely dense grid) \cite{FuchsBindingEnergy}.
%
%
\begin{figure}[t]
\centering
\includegraphics[scale=0.5]{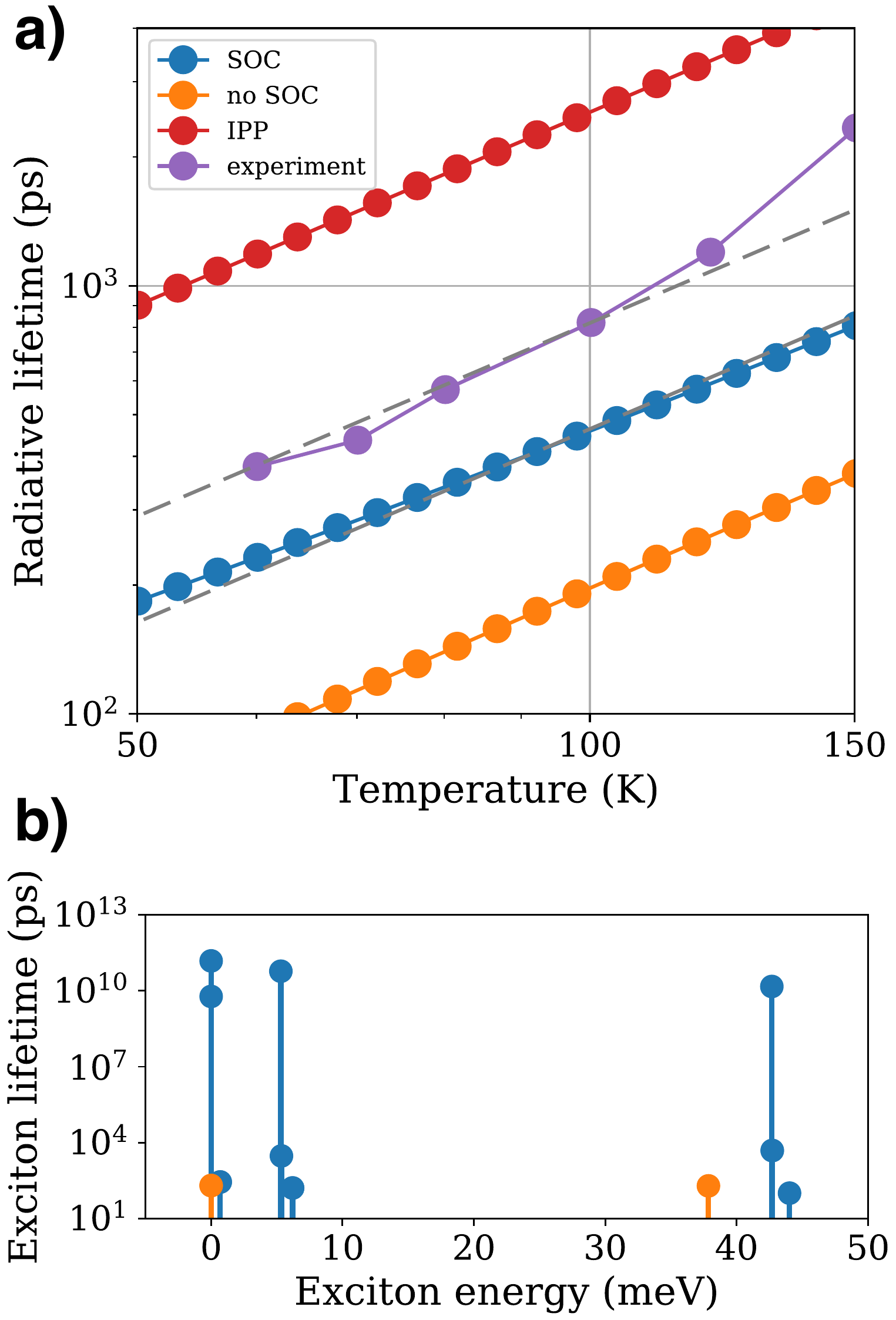}
\caption{a) Comparison of our radiative lifetimes computed by including (blue) or neglecting (orange) the SOC in the solution of the BSE, or obtained in the IPP by neglecting excitons (red). 
Experimental results from Ref.~\onlinecite{Im1997} (purple) are shown for comparison. 
The gray dashed lines show the $T^{3/2}$ trend predicted by our treatment at low temperature.  
b) The excitons contributing to the thermal average in Eq.~(\ref{eq:gammaeff}), along with their individual lifetimes at 100~K, computed with (blue) and without (orange) SOC. 
The zero of the energy axis is taken to be the lowest exciton energy for each case.
}
\label{Fig:lowT}
\end{figure}

Our computed radiative lifetimes between 50$-$150 K are shown in Fig.~\ref{Fig:lowT}a) along with experimental values from Ref.~\onlinecite{Im1997}, 
which are ideal for our comparison since they were measured in a relatively pure GaN crystal. 
At low temperatures up to 100~K, our first-principles radiative lifetimes, with SOC included, are of order 200$-$900~ps and are in very good agreement (within less than a factor of two) with experiment. 
We attribute the remaining discrepancy to small uncertainties in the computed exciton effective mass, transition dipoles, energies and occupations, 
plus inherent uncertainties in the experimental data.  
Both the computed and experimental lifetimes exhibit the intrinsic $T^{3/2}$ trend predicted by our approach~[see Eq.~(\ref{eq:gammatherm})]. 
As Fig.~\ref{Fig:lowT}a) shows, when neglecting excitons and using IPP transition dipoles and energies, 
one greatly overestimates the radiative lifetime. The IPP lifetimes are greater by nearly an order of magnitude compared to our treatment, 
which correctly includes excitons, and by over a factor of three compared to experiment.\\ 
\indent
As seen in Fig.~\ref{Fig:lowT}a), including SOC when computing the exciton states increases the radiative lifetimes by a factor of 2$-$3 and significantly improves the agreement with experiment.  
Though SOC is weak in GaN $-$ the valence band splitting at $\Gamma$ is only 5 meV in our calculations $-$ its inclusion is crucial for obtaining accurate exciton states.
Figure~\ref{Fig:lowT}b) shows the individual radiative lifetimes $\left<\gamma_S\right>^{-1}$ and relative energies of the low-energy excitons contributing to the thermal average in Eq.~(\ref{eq:gammaeff}), 
for both the cases where SOC is included and neglected. Without including spin and SOC, the exciton structure consists of three bright singlet excitons, two of which are degenerate. 
The lifetimes of all three excitons are nearly identical, and their value determines the radiative lifetime for the calculation without SOC. 
Including the SOC lifts the degeneracy of the two lowest bright excitons by $\sim$5 meV, and resolves the exciton fine structure, 
splitting each exciton into four states due to a doubling of the number of valence and conduction states that compose the electron-hole transitions. 
With SOC, we find dark excitons with lifetimes roughly 3$-$10 orders of magnitude longer than the excitons found without SOC. 
When included in the thermal average, these dark states are crucial as they increase the radiative lifetime compared to the average lifetime of the bright excitons alone.
The inclusion of SOC and the exciton fine structure are thus important for quantitative accuracy, even though SOC per se is weak in GaN.
Note that spin is always important. Even in the limit of vanishingly small SOC, the triplet states with ideally infinite lifetime would still suppress the average radiative rate in Eq.~(\ref{eq:gammaeff}) by a factor of 4, and thus increase the radiative lifetimes by the same factor compared to a calculation that does not include spin.
\\
\indent
%
%
Due to the small exciton binding energy in GaN, at high enough temperatures the excitons dissociate into free electrons and holes, 
which mainly recombine non-radiatively in GaN, giving rise to the lower radiative recombination rate and quantum yield seen experimentally above 100 K~\cite{Im1997,Yoon1996}. 
As a result of exciton dissociation, the measured radiative lifetime above $\sim$100~K increases more rapidly with temperature than the intrinsic $T^{3/2}$ trend~[see Fig.~\ref{Fig:lowT}a)]. 
We show a simple model to include exciton dissociation in our first-principles approach. 
Assuming that excitons and free carriers are in thermal equilibrium, we write the mass-action law for their concentrations as~\cite{Yoon1996}
\begin{equation} \label{eq:massaction1}
\frac{n_e n_h}{n_{\rm exc}} = \frac{[n_0 + \delta n]\delta p}{\delta n_{\rm exc}} = \kappa(T),
\end{equation}
where $n_e$, $n_h$, and $n_{\rm exc}$ are the electron, hole, and exciton densities, respectively, $n_0$ is the background electron density (from the doping), 
and $\delta n$, $\delta p$, and $\delta n_{\rm exc}$ are the excited electron, hole, and exciton densities, respectively, generated by an idealized optical pump or electrical current. 
The equilibrium constant $\kappa(T)$ is given by~\cite{Yoon1996} 
\begin{equation} \label{eq:massaction2}
\kappa(T) = 2\left(\frac{m_\text{red}k_BT}{2\pi\hbar^2}\right)^{3/2}e^{-E_b/k_BT},
\end{equation}
where $m_\text{red} \!=\! m_hm_e/(m_h+m_e)$ is the reduced mass of the exciton and $E_b$ its binding energy. 
We find a converged binding energy of 19.7 meV, in excellent agreement with the experimental value of 20.4 meV \cite{MuthGaNBE}, 
and we use a typical doping of $n_0 = 2.5\times 10^{16}$ $\text{cm}^{-3}$, taken from Ref.~\onlinecite{Im1997}.\\
\indent
Assuming that the relative recombination probability of free carriers and excitons is proportional to their concentration ratio, $P_\text{carr}/P_\text{exc} = \delta n / \delta n_{\rm exc}$, and using $P_\text{carr} + P_\text{exc} = 1$, 
we can obtain the probabilities for exciton and free carrier recombination. 
The measured radiative rate will be a weighted average of the rates of the two recombination processes, $\Gamma_\text{rad} = \Gamma_\text{carr}P_\text{carr} + \Gamma_\text{exc}P_\text{exc}$. 
We assume that $\Gamma_\text{carr}$ vanishes because free carriers recombine mainly via non-radiative channels, such as defect trapping, which is justified by the reported low quantum yield seen experimentally near room temperature \cite{Im1997}. 
The measured radiative rate due to excitons in equilibrium with carriers becomes $\Gamma_\text{rad} \approx \Gamma_\text{exc}/(1+\kappa(T)/n_0)$.
%
\begin{figure}[t]
\centering
\includegraphics[scale=0.55]{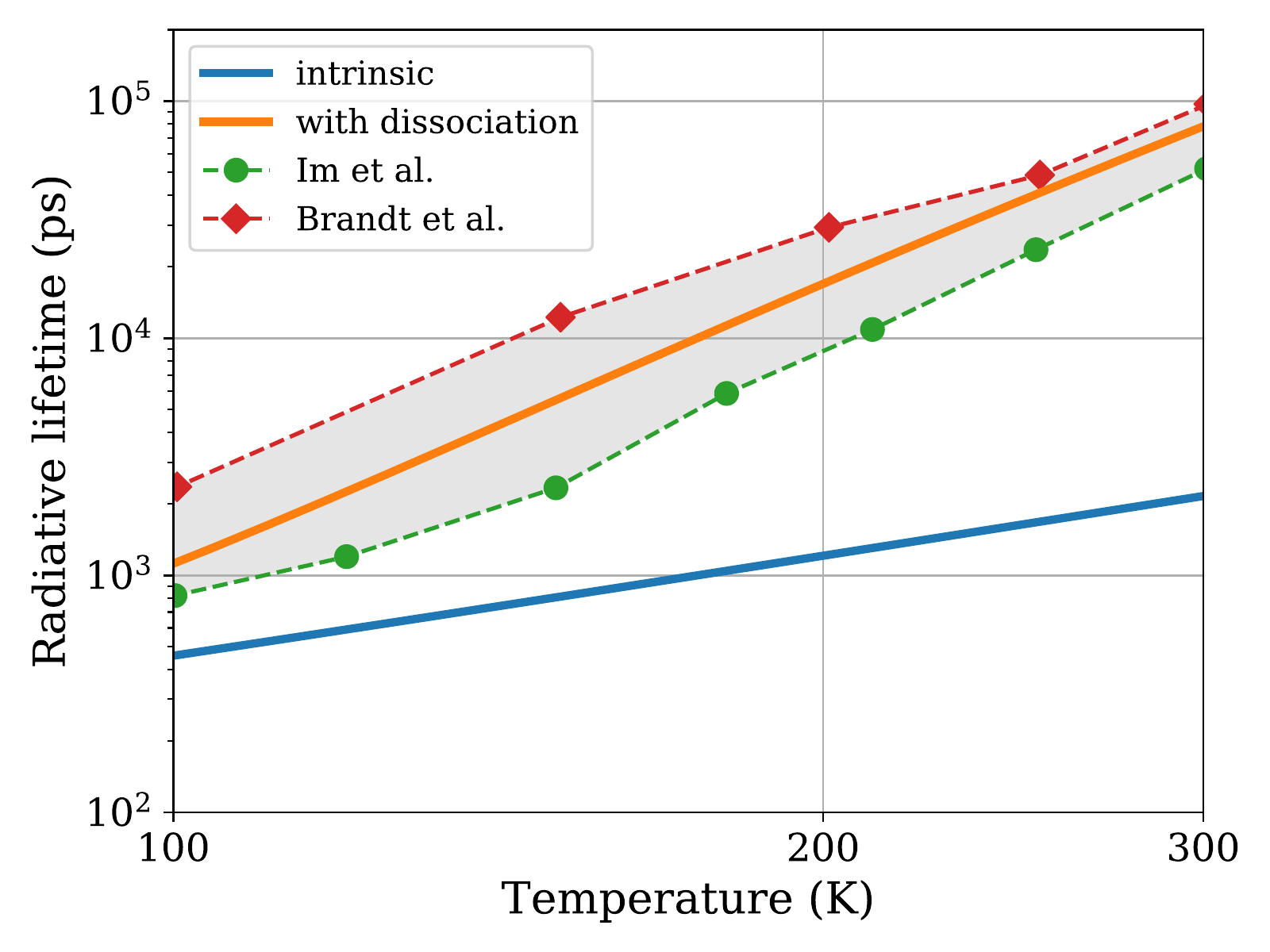}
\caption{Comparison of our computed radiative lifetimes including exciton dissociation (orange) above 100 K with experimental data from Refs.~\onlinecite{Im1997} and \onlinecite{Brandt1998}. 
Also shown is our computed intrinsic radiative lifetime (blue).}
\label{Fig:highT}
\end{figure}
Using this result, together with our computed effective masses and converged exciton binding energy, we are able to predict the exciton radiative lifetimes also above 100 K.\\
\indent
Figure~\ref{Fig:highT} compares the computed radiative lifetimes up to 300 K with experimental results taken from PL measurements in Refs.~\cite{Im1997, Brandt1998}. 
When thermal dissociation is included, the radiative lifetime agrees with experiment even in the 100$-$300~K temperature range, where the experimental data deviate from the intrinsic $T^{3/2}$ trend. 
Our ability to compute intrinsic exciton radiative lifetimes allows us to conclude that the radiative lifetime increase seen experimentally above 100 K is due to exciton thermal dissociation into free carriers. 
This conclusion is consistent with the results by Im et al.~\cite{Im1997}, who found that a similar exciton dissociation model could fit their experimental data at high temperature.\\
\indent
%
%
In summary, we developed accurate first-principles radiative lifetime calculations in GaN. Our method includes the electron-hole and spin-orbit interactions, thus correctly treating excitons and their fine structure. 
These advances allow us to compute intrinsic radiative lifetimes in very good agreement with experiment, and gain microscopic insight into the excitons associated with light emission in GaN. 
Future work will extend this analysis to nitride heterostructures and other solid state emitters such as InGaN. 
Our calculations can also be extended to nitride quantum wells, in which excitons are confined in two dimensions, by using modified exciton energies, dispersions, and transition dipoles. 
Our results add to the general framework we presented in Ref.~\onlinecite{Chen0D3D}, enabling precise predictions of radiative processes in solid-state light emitting materials, while shedding light on their ultrafast excited state dynamics. 
%
%

The authors thank Davide Sangalli for fruitful discussions. 
V.A.J. thanks the Resnick Sustainability Institute at Caltech for fellowship support. This work was partially supported by the Department of Energy under Grant No.~DE-SC0019166, which provided for theory and method development, 
and by the National Science Foundation under Grant No.~ACI-1642443, which provided for code development. This research used resources of the National Energy Research Scientific Computing Center, a DOE Office of Science User
Facility supported by the Office of Science of the U.S. Department of Energy under Contract No. DE-AC02-05CH11231. M.P. thanks CINECA for computational resources.


\begin{thebibliography}{35}%
\makeatletter
\providecommand \@ifxundefined [1]{%
 \@ifx{#1\undefined}
}%
\providecommand \@ifnum [1]{%
 \ifnum #1\expandafter \@firstoftwo
 \else \expandafter \@secondoftwo
 \fi
}%
\providecommand \@ifx [1]{%
 \ifx #1\expandafter \@firstoftwo
 \else \expandafter \@secondoftwo
 \fi
}%
\providecommand \natexlab [1]{#1}%
\providecommand \enquote  [1]{``#1''}%
\providecommand \bibnamefont  [1]{#1}%
\providecommand \bibfnamefont [1]{#1}%
\providecommand \citenamefont [1]{#1}%
\providecommand \href@noop [0]{\@secondoftwo}%
\providecommand \href [0]{\begingroup \@sanitize@url \@href}%
\providecommand \@href[1]{\@@startlink{#1}\@@href}%
\providecommand \@@href[1]{\endgroup#1\@@endlink}%
\providecommand \@sanitize@url [0]{\catcode `\\12\catcode `\$12\catcode
  `\&12\catcode `\#12\catcode `\^12\catcode `\_12\catcode `\%12\relax}%
\providecommand \@@startlink[1]{}%
\providecommand \@@endlink[0]{}%
\providecommand \url  [0]{\begingroup\@sanitize@url \@url }%
\providecommand \@url [1]{\endgroup\@href {#1}{\urlprefix }}%
\providecommand \urlprefix  [0]{URL }%
\providecommand \Eprint [0]{\href }%
\providecommand \doibase [0]{https://doi.org/}%
\providecommand \selectlanguage [0]{\@gobble}%
\providecommand \bibinfo  [0]{\@secondoftwo}%
\providecommand \bibfield  [0]{\@secondoftwo}%
\providecommand \translation [1]{[#1]}%
\providecommand \BibitemOpen [0]{}%
\providecommand \bibitemStop [0]{}%
\providecommand \bibitemNoStop [0]{.\EOS\space}%
\providecommand \EOS [0]{\spacefactor3000\relax}%
\providecommand \BibitemShut  [1]{\csname bibitem#1\endcsname}%
\let\auto@bib@innerbib\@empty
\bibitem [{\citenamefont {Knox}(1963)}]{knox1963}%
  \BibitemOpen
  \bibfield  {author} {\bibinfo {author} {\bibfnamefont {R.~S.}\ \bibnamefont
  {Knox}},\ }\href@noop {} {\emph {\bibinfo {title} {Theory of Excitons (Solid
  State Phys. Suppl. 5)}}},\ Vol.~\bibinfo {volume} {5}\ (\bibinfo  {publisher}
  {Academic Press, New York},\ \bibinfo {year} {1963})\BibitemShut {NoStop}%
\bibitem [{\citenamefont {Pimputkar}\ \emph {et~al.}(2009)\citenamefont
  {Pimputkar}, \citenamefont {Speck}, \citenamefont {DenBaars},\ and\
  \citenamefont {Nakamura}}]{LEDprospects}%
  \BibitemOpen
  \bibfield  {author} {\bibinfo {author} {\bibfnamefont {S.}~\bibnamefont
  {Pimputkar}}, \bibinfo {author} {\bibfnamefont {J.~S.}\ \bibnamefont
  {Speck}}, \bibinfo {author} {\bibfnamefont {S.~P.}\ \bibnamefont
  {DenBaars}},\ and\ \bibinfo {author} {\bibfnamefont {S.}~\bibnamefont
  {Nakamura}},\ }\bibfield  {title} {\bibinfo {title} {Prospects for {LED}
  lighting},\ }\href {http://dx.doi.org/10.1038/nphoton.2009.32} {\bibfield
  {journal} {\bibinfo  {journal} {Nat. Photonics}\ }\textbf {\bibinfo {volume}
  {3}},\ \bibinfo {pages} {180} (\bibinfo {year} {2009})}\BibitemShut {NoStop}%
\bibitem [{\citenamefont {Muth}\ \emph {et~al.}(1997)\citenamefont {Muth},
  \citenamefont {Lee}, \citenamefont {Shmagin}, \citenamefont {Kolbas},
  \citenamefont {Casey}, \citenamefont {Keller}, \citenamefont {Mishra},\ and\
  \citenamefont {DenBaars}}]{MuthGaNBE}%
  \BibitemOpen
  \bibfield  {author} {\bibinfo {author} {\bibfnamefont {J.~F.}\ \bibnamefont
  {Muth}}, \bibinfo {author} {\bibfnamefont {J.~H.}\ \bibnamefont {Lee}},
  \bibinfo {author} {\bibfnamefont {I.~K.}\ \bibnamefont {Shmagin}}, \bibinfo
  {author} {\bibfnamefont {R.~M.}\ \bibnamefont {Kolbas}}, \bibinfo {author}
  {\bibfnamefont {H.~C.}\ \bibnamefont {Casey}}, \bibinfo {author}
  {\bibfnamefont {B.~P.}\ \bibnamefont {Keller}}, \bibinfo {author}
  {\bibfnamefont {U.~K.}\ \bibnamefont {Mishra}},\ and\ \bibinfo {author}
  {\bibfnamefont {S.~P.}\ \bibnamefont {DenBaars}},\ }\bibfield  {title}
  {\bibinfo {title} {Absorption coefficient, energy gap, exciton binding
  energy, and recombination lifetime of {GaN} obtained from transmission
  measurements},\ }\href {https://doi.org/10.1063/1.120191} {\bibfield
  {journal} {\bibinfo  {journal} {Appl. Phys. Lett.}\ }\textbf {\bibinfo
  {volume} {71}},\ \bibinfo {pages} {2572--2574} (\bibinfo {year}
  {1997})}\BibitemShut {NoStop}%
\bibitem [{\citenamefont {Laskowski}\ \emph {et~al.}(2005)\citenamefont
  {Laskowski}, \citenamefont {Christensen}, \citenamefont {Santi},\ and\
  \citenamefont {Ambrosch-Draxl}}]{Laskowski2005}%
  \BibitemOpen
  \bibfield  {author} {\bibinfo {author} {\bibfnamefont {R.}~\bibnamefont
  {Laskowski}}, \bibinfo {author} {\bibfnamefont {N.~E.}\ \bibnamefont
  {Christensen}}, \bibinfo {author} {\bibfnamefont {G.}~\bibnamefont {Santi}},\
  and\ \bibinfo {author} {\bibfnamefont {C.}~\bibnamefont {Ambrosch-Draxl}},\
  }\bibfield  {title} {\bibinfo {title} {Ab initio calculations of excitons in
  {GaN}},\ }\href {https://doi.org/10.1103/PhysRevB.72.035204} {\bibfield
  {journal} {\bibinfo  {journal} {Phys. Rev. B}\ }\textbf {\bibinfo {volume}
  {72}},\ \bibinfo {pages} {035204} (\bibinfo {year} {2005})}\BibitemShut
  {NoStop}%
\bibitem [{\citenamefont {Harris}\ \emph {et~al.}(1995)\citenamefont {Harris},
  \citenamefont {Monemar}, \citenamefont {Amano},\ and\ \citenamefont
  {Akasaki}}]{Harris1995}%
  \BibitemOpen
  \bibfield  {author} {\bibinfo {author} {\bibfnamefont {C.~I.}\ \bibnamefont
  {Harris}}, \bibinfo {author} {\bibfnamefont {B.}~\bibnamefont {Monemar}},
  \bibinfo {author} {\bibfnamefont {H.}~\bibnamefont {Amano}},\ and\ \bibinfo
  {author} {\bibfnamefont {I.}~\bibnamefont {Akasaki}},\ }\bibfield  {title}
  {\bibinfo {title} {Exciton lifetimes in {GaN} and {GaInN}},\ }\href
  {https://doi.org/10.1063/1.115522} {\bibfield  {journal} {\bibinfo  {journal}
  {Appl. Phys. Lett.}\ }\textbf {\bibinfo {volume} {67}},\ \bibinfo {pages}
  {840--842} (\bibinfo {year} {1995})}\BibitemShut {NoStop}%
\bibitem [{\citenamefont {Chen}\ \emph {et~al.}(1996)\citenamefont {Chen},
  \citenamefont {Smith}, \citenamefont {Lin}, \citenamefont {Jiang},
  \citenamefont {Wei}, \citenamefont {Asif~Khan},\ and\ \citenamefont
  {Sun}}]{Chen1996}%
  \BibitemOpen
  \bibfield  {author} {\bibinfo {author} {\bibfnamefont {G.}~\bibnamefont
  {Chen}}, \bibinfo {author} {\bibfnamefont {M.}~\bibnamefont {Smith}},
  \bibinfo {author} {\bibfnamefont {J.}~\bibnamefont {Lin}}, \bibinfo {author}
  {\bibfnamefont {H.}~\bibnamefont {Jiang}}, \bibinfo {author} {\bibfnamefont
  {S.-H.}\ \bibnamefont {Wei}}, \bibinfo {author} {\bibfnamefont
  {M.}~\bibnamefont {Asif~Khan}},\ and\ \bibinfo {author} {\bibfnamefont
  {C.}~\bibnamefont {Sun}},\ }\bibfield  {title} {\bibinfo {title} {Fundamental
  optical transitions in {GaN}},\ }\href
  {https://doi.org/doi.org/10.1063/1.116606} {\bibfield  {journal} {\bibinfo
  {journal} {Appl. Phys. Lett.}\ }\textbf {\bibinfo {volume} {68}},\ \bibinfo
  {pages} {2784--2786} (\bibinfo {year} {1996})}\BibitemShut {NoStop}%
\bibitem [{\citenamefont {Im}\ \emph {et~al.}(1997)\citenamefont {Im},
  \citenamefont {Moritz}, \citenamefont {Steuber}, \citenamefont {Härle},
  \citenamefont {Scholz},\ and\ \citenamefont {Hangleiter}}]{Im1997}%
  \BibitemOpen
  \bibfield  {author} {\bibinfo {author} {\bibfnamefont {J.~S.}\ \bibnamefont
  {Im}}, \bibinfo {author} {\bibfnamefont {A.}~\bibnamefont {Moritz}}, \bibinfo
  {author} {\bibfnamefont {F.}~\bibnamefont {Steuber}}, \bibinfo {author}
  {\bibfnamefont {V.}~\bibnamefont {Härle}}, \bibinfo {author} {\bibfnamefont
  {F.}~\bibnamefont {Scholz}},\ and\ \bibinfo {author} {\bibfnamefont
  {A.}~\bibnamefont {Hangleiter}},\ }\bibfield  {title} {\bibinfo {title}
  {Radiative carrier lifetime, momentum matrix element, and hole effective mass
  in {GaN}},\ }\href {https://doi.org/10.1063/1.118293} {\bibfield  {journal}
  {\bibinfo  {journal} {Appl. Phys. Lett.}\ }\textbf {\bibinfo {volume} {70}},\
  \bibinfo {pages} {631--633} (\bibinfo {year} {1997})}\BibitemShut {NoStop}%
\bibitem [{\citenamefont {Brandt}\ \emph {et~al.}(1998)\citenamefont {Brandt},
  \citenamefont {Ringling}, \citenamefont {Ploog}, \citenamefont {W\"unsche},\
  and\ \citenamefont {Henneberger}}]{Brandt1998}%
  \BibitemOpen
  \bibfield  {author} {\bibinfo {author} {\bibfnamefont {O.}~\bibnamefont
  {Brandt}}, \bibinfo {author} {\bibfnamefont {J.}~\bibnamefont {Ringling}},
  \bibinfo {author} {\bibfnamefont {K.~H.}\ \bibnamefont {Ploog}}, \bibinfo
  {author} {\bibfnamefont {H.-J.}\ \bibnamefont {W\"unsche}},\ and\ \bibinfo
  {author} {\bibfnamefont {F.}~\bibnamefont {Henneberger}},\ }\bibfield
  {title} {\bibinfo {title} {Temperature dependence of the radiative lifetime
  in {GaN}},\ }\href {https://doi.org/10.1103/PhysRevB.58.R15977} {\bibfield
  {journal} {\bibinfo  {journal} {Phys. Rev. B}\ }\textbf {\bibinfo {volume}
  {58}},\ \bibinfo {pages} {R15977--R15980} (\bibinfo {year}
  {1998})}\BibitemShut {NoStop}%
\bibitem [{\citenamefont {Jhalani}\ \emph {et~al.}(2017)\citenamefont
  {Jhalani}, \citenamefont {Zhou},\ and\ \citenamefont
  {Bernardi}}]{Vatsal2017}%
  \BibitemOpen
  \bibfield  {author} {\bibinfo {author} {\bibfnamefont {V.~A.}\ \bibnamefont
  {Jhalani}}, \bibinfo {author} {\bibfnamefont {J.-J.}\ \bibnamefont {Zhou}},\
  and\ \bibinfo {author} {\bibfnamefont {M.}~\bibnamefont {Bernardi}},\
  }\bibfield  {title} {\bibinfo {title} {Ultrafast hot carrier dynamics in
  {GaN} and its impact on the efficiency droop},\ }\href
  {https://doi.org/10.1021/acs.nanolett.7b02212} {\bibfield  {journal}
  {\bibinfo  {journal} {Nano Lett.}\ }\textbf {\bibinfo {volume} {17}},\
  \bibinfo {pages} {5012--5019} (\bibinfo {year} {2017})}\BibitemShut {NoStop}%
\bibitem [{\citenamefont {Wolfe}(1982)}]{WolfeExcTherm}%
  \BibitemOpen
  \bibfield  {author} {\bibinfo {author} {\bibfnamefont {J.~P.}\ \bibnamefont
  {Wolfe}},\ }\bibfield  {title} {\bibinfo {title} {Thermodynamics of excitons
  in semiconductors},\ }\href {https://doi.org/10.1063/1.2914968} {\bibfield
  {journal} {\bibinfo  {journal} {Phys. Today}\ }\textbf {\bibinfo {volume}
  {35}},\ \bibinfo {pages} {46--54} (\bibinfo {year} {1982})}\BibitemShut
  {NoStop}%
\bibitem [{\citenamefont {Ridley}(2013)}]{Ridley}%
  \BibitemOpen
  \bibfield  {author} {\bibinfo {author} {\bibfnamefont {B.~K.}\ \bibnamefont
  {Ridley}},\ }\href@noop {} {\emph {\bibinfo {title} {Quantum Processes in
  Semiconductors}}}\ (\bibinfo  {publisher} {Oxford University Press},\
  \bibinfo {year} {2013})\BibitemShut {NoStop}%
\bibitem [{\citenamefont {Zhang}\ \emph {et~al.}(2018)\citenamefont {Zhang},
  \citenamefont {Shen}, \citenamefont {Wang},\ and\ \citenamefont {Van~de
  Walle}}]{zhang2018perov}%
  \BibitemOpen
  \bibfield  {author} {\bibinfo {author} {\bibfnamefont {X.}~\bibnamefont
  {Zhang}}, \bibinfo {author} {\bibfnamefont {J.-X.}\ \bibnamefont {Shen}},
  \bibinfo {author} {\bibfnamefont {W.}~\bibnamefont {Wang}},\ and\ \bibinfo
  {author} {\bibfnamefont {C.~G.}\ \bibnamefont {Van~de Walle}},\ }\bibfield
  {title} {\bibinfo {title} {First-principles analysis of radiative
  recombination in lead-halide perovskites},\ }\href
  {https://doi.org/10.1021/acsenergylett.8b01297} {\bibfield  {journal}
  {\bibinfo  {journal} {ACS Energy Lett.}\ }\textbf {\bibinfo {volume} {3}},\
  \bibinfo {pages} {2329--2334} (\bibinfo {year} {2018})}\BibitemShut {NoStop}%
\bibitem [{\citenamefont {Kioupakis}\ \emph {et~al.}(2013)\citenamefont
  {Kioupakis}, \citenamefont {Yan}, \citenamefont {Steiauf},\ and\
  \citenamefont {Van~de Walle}}]{kioupakis2013nitride}%
  \BibitemOpen
  \bibfield  {author} {\bibinfo {author} {\bibfnamefont {E.}~\bibnamefont
  {Kioupakis}}, \bibinfo {author} {\bibfnamefont {Q.}~\bibnamefont {Yan}},
  \bibinfo {author} {\bibfnamefont {D.}~\bibnamefont {Steiauf}},\ and\ \bibinfo
  {author} {\bibfnamefont {C.~G.}\ \bibnamefont {Van~de Walle}},\ }\bibfield
  {title} {\bibinfo {title} {Temperature and carrier-density dependence of
  auger and radiative recombination in nitride optoelectronic devices},\ }\href
  {https://doi.org/10.1088/1367-2630/15/12/125006} {\bibfield  {journal}
  {\bibinfo  {journal} {New J. Phys.}\ }\textbf {\bibinfo {volume} {15}},\
  \bibinfo {pages} {125006} (\bibinfo {year} {2013})}\BibitemShut {NoStop}%
\bibitem [{\citenamefont {Chen}\ \emph {et~al.}(2019)\citenamefont {Chen},
  \citenamefont {Jhalani}, \citenamefont {Palummo},\ and\ \citenamefont
  {Bernardi}}]{Chen0D3D}%
  \BibitemOpen
  \bibfield  {author} {\bibinfo {author} {\bibfnamefont {H.-Y.}\ \bibnamefont
  {Chen}}, \bibinfo {author} {\bibfnamefont {V.~A.}\ \bibnamefont {Jhalani}},
  \bibinfo {author} {\bibfnamefont {M.}~\bibnamefont {Palummo}},\ and\ \bibinfo
  {author} {\bibfnamefont {M.}~\bibnamefont {Bernardi}},\ }\bibfield  {title}
  {\bibinfo {title} {Ab initio calculations of exciton radiative lifetimes in
  bulk crystals, nanostructures, and molecules},\ }\href
  {https://doi.org/10.1103/PhysRevB.100.075135} {\bibfield  {journal} {\bibinfo
   {journal} {Phys. Rev. B}\ }\textbf {\bibinfo {volume} {100}},\ \bibinfo
  {pages} {075135} (\bibinfo {year} {2019})}\BibitemShut {NoStop}%
\bibitem [{\citenamefont {Spataru}\ \emph {et~al.}(2005)\citenamefont
  {Spataru}, \citenamefont {Ismail-Beigi}, \citenamefont {Capaz},\ and\
  \citenamefont {Louie}}]{Spataru2005}%
  \BibitemOpen
  \bibfield  {author} {\bibinfo {author} {\bibfnamefont {C.~D.}\ \bibnamefont
  {Spataru}}, \bibinfo {author} {\bibfnamefont {S.}~\bibnamefont
  {Ismail-Beigi}}, \bibinfo {author} {\bibfnamefont {R.~B.}\ \bibnamefont
  {Capaz}},\ and\ \bibinfo {author} {\bibfnamefont {S.~G.}\ \bibnamefont
  {Louie}},\ }\bibfield  {title} {\bibinfo {title} {Theory and ab initio
  calculation of radiative lifetime of excitons in semiconducting carbon
  nanotubes},\ }\href {https://doi.org/10.1103/PhysRevLett.95.247402}
  {\bibfield  {journal} {\bibinfo  {journal} {Phys. Rev. Lett.}\ }\textbf
  {\bibinfo {volume} {95}},\ \bibinfo {pages} {247402} (\bibinfo {year}
  {2005})}\BibitemShut {NoStop}%
\bibitem [{\citenamefont {Palummo}\ \emph {et~al.}(2015)\citenamefont
  {Palummo}, \citenamefont {Bernardi},\ and\ \citenamefont
  {Grossman}}]{PalummoTMDC}%
  \BibitemOpen
  \bibfield  {author} {\bibinfo {author} {\bibfnamefont {M.}~\bibnamefont
  {Palummo}}, \bibinfo {author} {\bibfnamefont {M.}~\bibnamefont {Bernardi}},\
  and\ \bibinfo {author} {\bibfnamefont {J.~C.}\ \bibnamefont {Grossman}},\
  }\bibfield  {title} {\bibinfo {title} {Exciton radiative lifetimes in
  two-dimensional transition metal dichalcogenides},\ }\href
  {https://doi.org/10.1021/nl503799t} {\bibfield  {journal} {\bibinfo
  {journal} {Nano Lett.}\ }\textbf {\bibinfo {volume} {15}},\ \bibinfo {pages}
  {2794--2800} (\bibinfo {year} {2015})}\BibitemShut {NoStop}%
\bibitem [{\citenamefont {Chen}\ \emph {et~al.}(2018)\citenamefont {Chen},
  \citenamefont {Palummo}, \citenamefont {Sangalli},\ and\ \citenamefont
  {Bernardi}}]{ChenTMDC}%
  \BibitemOpen
  \bibfield  {author} {\bibinfo {author} {\bibfnamefont {H.-Y.}\ \bibnamefont
  {Chen}}, \bibinfo {author} {\bibfnamefont {M.}~\bibnamefont {Palummo}},
  \bibinfo {author} {\bibfnamefont {D.}~\bibnamefont {Sangalli}},\ and\
  \bibinfo {author} {\bibfnamefont {M.}~\bibnamefont {Bernardi}},\ }\bibfield
  {title} {\bibinfo {title} {Theory and ab initio computation of the
  anisotropic light emission in monolayer transition metal dichalcogenides},\
  }\href {https://doi.org/10.1021/acs.nanolett.8b01114} {\bibfield  {journal}
  {\bibinfo  {journal} {Nano Lett.}\ }\textbf {\bibinfo {volume} {18}},\
  \bibinfo {pages} {3839--3843} (\bibinfo {year} {2018})}\BibitemShut {NoStop}%
\bibitem [{\citenamefont {Gao}\ \emph {et~al.}(2017)\citenamefont {Gao},
  \citenamefont {Yang},\ and\ \citenamefont {Spataru}}]{GaoTMDC}%
  \BibitemOpen
  \bibfield  {author} {\bibinfo {author} {\bibfnamefont {S.}~\bibnamefont
  {Gao}}, \bibinfo {author} {\bibfnamefont {L.}~\bibnamefont {Yang}},\ and\
  \bibinfo {author} {\bibfnamefont {C.~D.}\ \bibnamefont {Spataru}},\
  }\bibfield  {title} {\bibinfo {title} {Interlayer coupling and gate-tunable
  excitons in transition metal dichalcogenide heterostructures},\ }\href
  {https://doi.org/10.1021/acs.nanolett.7b04021} {\bibfield  {journal}
  {\bibinfo  {journal} {Nano Lett.}\ }\textbf {\bibinfo {volume} {17}},\
  \bibinfo {pages} {7809--7813} (\bibinfo {year} {2017})}\BibitemShut {NoStop}%
\bibitem [{\citenamefont {Strinati}(1984)}]{Strinati1984}%
  \BibitemOpen
  \bibfield  {author} {\bibinfo {author} {\bibfnamefont {G.}~\bibnamefont
  {Strinati}},\ }\bibfield  {title} {\bibinfo {title} {Effects of dynamical
  screening on resonances at inner-shell thresholds in semiconductors},\ }\href
  {https://doi.org/10.1103/PhysRevB.29.5718} {\bibfield  {journal} {\bibinfo
  {journal} {Phys. Rev. B}\ }\textbf {\bibinfo {volume} {29}},\ \bibinfo
  {pages} {5718--5726} (\bibinfo {year} {1984})}\BibitemShut {NoStop}%
\bibitem [{\citenamefont {Rohlfing}\ and\ \citenamefont
  {Louie}(2000)}]{RohlfingBSE}%
  \BibitemOpen
  \bibfield  {author} {\bibinfo {author} {\bibfnamefont {M.}~\bibnamefont
  {Rohlfing}}\ and\ \bibinfo {author} {\bibfnamefont {S.~G.}\ \bibnamefont
  {Louie}},\ }\bibfield  {title} {\bibinfo {title} {Electron-hole excitations
  and optical spectra from first principles},\ }\href
  {https://doi.org/10.1103/PhysRevB.62.4927} {\bibfield  {journal} {\bibinfo
  {journal} {Phys. Rev. B}\ }\textbf {\bibinfo {volume} {62}},\ \bibinfo
  {pages} {4927--4944} (\bibinfo {year} {2000})}\BibitemShut {NoStop}%
\bibitem [{\citenamefont {Glauber}\ and\ \citenamefont
  {Lewenstein}(1991)}]{Glauber1991}%
  \BibitemOpen
  \bibfield  {author} {\bibinfo {author} {\bibfnamefont {R.~J.}\ \bibnamefont
  {Glauber}}\ and\ \bibinfo {author} {\bibfnamefont {M.}~\bibnamefont
  {Lewenstein}},\ }\bibfield  {title} {\bibinfo {title} {Quantum optics of
  dielectric media},\ }\href {https://doi.org/10.1103/PhysRevA.43.467}
  {\bibfield  {journal} {\bibinfo  {journal} {Phys. Rev. A}\ }\textbf {\bibinfo
  {volume} {43}},\ \bibinfo {pages} {467--491} (\bibinfo {year}
  {1991})}\BibitemShut {NoStop}%
\bibitem [{\citenamefont {Sangalli}\ \emph {et~al.}(2017)\citenamefont
  {Sangalli}, \citenamefont {Berger}, \citenamefont {Attaccalite},
  \citenamefont {Gr\"uning},\ and\ \citenamefont {Romaniello}}]{Sangalli2017}%
  \BibitemOpen
  \bibfield  {author} {\bibinfo {author} {\bibfnamefont {D.}~\bibnamefont
  {Sangalli}}, \bibinfo {author} {\bibfnamefont {J.~A.}\ \bibnamefont
  {Berger}}, \bibinfo {author} {\bibfnamefont {C.}~\bibnamefont {Attaccalite}},
  \bibinfo {author} {\bibfnamefont {M.}~\bibnamefont {Gr\"uning}},\ and\
  \bibinfo {author} {\bibfnamefont {P.}~\bibnamefont {Romaniello}},\ }\bibfield
   {title} {\bibinfo {title} {Optical properties of periodic systems within the
  current-current response framework: Pitfalls and remedies},\ }\href
  {https://doi.org/10.1103/PhysRevB.95.155203} {\bibfield  {journal} {\bibinfo
  {journal} {Phys. Rev. B}\ }\textbf {\bibinfo {volume} {95}},\ \bibinfo
  {pages} {155203} (\bibinfo {year} {2017})}\BibitemShut {NoStop}%
\bibitem [{\citenamefont {Perdew}\ \emph {et~al.}(1996)\citenamefont {Perdew},
  \citenamefont {Burke},\ and\ \citenamefont {Ernzerhof}}]{GGA}%
  \BibitemOpen
  \bibfield  {author} {\bibinfo {author} {\bibfnamefont {J.~P.}\ \bibnamefont
  {Perdew}}, \bibinfo {author} {\bibfnamefont {K.}~\bibnamefont {Burke}},\ and\
  \bibinfo {author} {\bibfnamefont {M.}~\bibnamefont {Ernzerhof}},\ }\bibfield
  {title} {\bibinfo {title} {Generalized gradient approximation made simple},\
  }\href {https://doi.org/10.1103/PhysRevLett.77.3865} {\bibfield  {journal}
  {\bibinfo  {journal} {Phys. Rev. Lett.}\ }\textbf {\bibinfo {volume} {77}},\
  \bibinfo {pages} {3865--3868} (\bibinfo {year} {1996})}\BibitemShut {NoStop}%
\bibitem [{\citenamefont {Perdew}\ \emph {et~al.}(2008)\citenamefont {Perdew},
  \citenamefont {Ruzsinszky}, \citenamefont {Csonka}, \citenamefont {Vydrov},
  \citenamefont {Scuseria}, \citenamefont {Constantin}, \citenamefont {Zhou},\
  and\ \citenamefont {Burke}}]{PBEsol}%
  \BibitemOpen
  \bibfield  {author} {\bibinfo {author} {\bibfnamefont {J.~P.}\ \bibnamefont
  {Perdew}}, \bibinfo {author} {\bibfnamefont {A.}~\bibnamefont {Ruzsinszky}},
  \bibinfo {author} {\bibfnamefont {G.~I.}\ \bibnamefont {Csonka}}, \bibinfo
  {author} {\bibfnamefont {O.~A.}\ \bibnamefont {Vydrov}}, \bibinfo {author}
  {\bibfnamefont {G.~E.}\ \bibnamefont {Scuseria}}, \bibinfo {author}
  {\bibfnamefont {L.~A.}\ \bibnamefont {Constantin}}, \bibinfo {author}
  {\bibfnamefont {X.}~\bibnamefont {Zhou}},\ and\ \bibinfo {author}
  {\bibfnamefont {K.}~\bibnamefont {Burke}},\ }\bibfield  {title} {\bibinfo
  {title} {Restoring the density-gradient expansion for exchange in solids and
  surfaces},\ }\href {https://doi.org/10.1103/PhysRevLett.100.136406}
  {\bibfield  {journal} {\bibinfo  {journal} {Phys. Rev. Lett.}\ }\textbf
  {\bibinfo {volume} {100}},\ \bibinfo {pages} {136406} (\bibinfo {year}
  {2008})}\BibitemShut {NoStop}%
\bibitem [{\citenamefont {Giannozzi}\ \emph {et~al.}(2009)\citenamefont
  {Giannozzi}, \citenamefont {Baroni}, \citenamefont {Bonini}, \citenamefont
  {Calandra}, \citenamefont {Car}, \citenamefont {Cavazzoni}, \citenamefont
  {Ceresoli}, \citenamefont {Chiarotti}, \citenamefont {Cococcioni},
  \citenamefont {Dabo} \emph {et~al.}}]{QE}%
  \BibitemOpen
  \bibfield  {author} {\bibinfo {author} {\bibfnamefont {P.}~\bibnamefont
  {Giannozzi}}, \bibinfo {author} {\bibfnamefont {S.}~\bibnamefont {Baroni}},
  \bibinfo {author} {\bibfnamefont {N.}~\bibnamefont {Bonini}}, \bibinfo
  {author} {\bibfnamefont {M.}~\bibnamefont {Calandra}}, \bibinfo {author}
  {\bibfnamefont {R.}~\bibnamefont {Car}}, \bibinfo {author} {\bibfnamefont
  {C.}~\bibnamefont {Cavazzoni}}, \bibinfo {author} {\bibfnamefont
  {D.}~\bibnamefont {Ceresoli}}, \bibinfo {author} {\bibfnamefont {G.~L.}\
  \bibnamefont {Chiarotti}}, \bibinfo {author} {\bibfnamefont {M.}~\bibnamefont
  {Cococcioni}}, \bibinfo {author} {\bibfnamefont {I.}~\bibnamefont {Dabo}},
  \emph {et~al.},\ }\bibfield  {title} {\bibinfo {title} {{QUANTUM ESPRESSO}: a
  modular and open-source software project for quantum simulations of
  materials},\ }\href {http://www.quantum-espresso.org} {\bibfield  {journal}
  {\bibinfo  {journal} {J. Phys.: Condens. Matter}\ }\textbf {\bibinfo {volume}
  {21}},\ \bibinfo {pages} {395502} (\bibinfo {year} {2009})}\BibitemShut
  {NoStop}%
\bibitem [{\citenamefont {Hamann}(2013)}]{ONCVPSP}%
  \BibitemOpen
  \bibfield  {author} {\bibinfo {author} {\bibfnamefont {D.~R.}\ \bibnamefont
  {Hamann}},\ }\bibfield  {title} {\bibinfo {title} {Optimized norm-conserving
  {Vanderbilt} pseudopotentials},\ }\href
  {https://doi.org/10.1103/PhysRevB.88.085117} {\bibfield  {journal} {\bibinfo
  {journal} {Phys. Rev. B}\ }\textbf {\bibinfo {volume} {88}},\ \bibinfo
  {pages} {085117} (\bibinfo {year} {2013})}\BibitemShut {NoStop}%
\bibitem [{\citenamefont {van Setten}\ \emph {et~al.}(2018)\citenamefont {van
  Setten}, \citenamefont {Giantomassi}, \citenamefont {Bousquet}, \citenamefont
  {Verstraete}, \citenamefont {Hamann}, \citenamefont {Gonze},\ and\
  \citenamefont {Rignanese}}]{Dojo}%
  \BibitemOpen
  \bibfield  {author} {\bibinfo {author} {\bibfnamefont {M.}~\bibnamefont {van
  Setten}}, \bibinfo {author} {\bibfnamefont {M.}~\bibnamefont {Giantomassi}},
  \bibinfo {author} {\bibfnamefont {E.}~\bibnamefont {Bousquet}}, \bibinfo
  {author} {\bibfnamefont {M.}~\bibnamefont {Verstraete}}, \bibinfo {author}
  {\bibfnamefont {D.}~\bibnamefont {Hamann}}, \bibinfo {author} {\bibfnamefont
  {X.}~\bibnamefont {Gonze}},\ and\ \bibinfo {author} {\bibfnamefont {G.-M.}\
  \bibnamefont {Rignanese}},\ }\bibfield  {title} {\bibinfo {title} {The
  {PseudoDojo}: {Training} and grading a 85 element optimized norm-conserving
  pseudopotential table},\ }\href
  {http://www.sciencedirect.com/science/article/pii/S0010465518300250}
  {\bibfield  {journal} {\bibinfo  {journal} {Comput. Phys. Commun.}\ }\textbf
  {\bibinfo {volume} {226}},\ \bibinfo {pages} {39 -- 54} (\bibinfo {year}
  {2018})}\BibitemShut {NoStop}%
\bibitem [{\citenamefont {Louie}\ \emph {et~al.}(1982)\citenamefont {Louie},
  \citenamefont {Froyen},\ and\ \citenamefont {Cohen}}]{NLCC}%
  \BibitemOpen
  \bibfield  {author} {\bibinfo {author} {\bibfnamefont {S.~G.}\ \bibnamefont
  {Louie}}, \bibinfo {author} {\bibfnamefont {S.}~\bibnamefont {Froyen}},\ and\
  \bibinfo {author} {\bibfnamefont {M.~L.}\ \bibnamefont {Cohen}},\ }\bibfield
  {title} {\bibinfo {title} {Nonlinear ionic pseudopotentials in
  spin-density-functional calculations},\ }\href
  {https://doi.org/10.1103/PhysRevB.26.1738} {\bibfield  {journal} {\bibinfo
  {journal} {Phys. Rev. B}\ }\textbf {\bibinfo {volume} {26}},\ \bibinfo
  {pages} {1738--1742} (\bibinfo {year} {1982})}\BibitemShut {NoStop}%
\bibitem [{\citenamefont {Rubio}\ \emph {et~al.}(1993)\citenamefont {Rubio},
  \citenamefont {Corkill}, \citenamefont {Cohen}, \citenamefont {Shirley},\
  and\ \citenamefont {Louie}}]{RubioGaN}%
  \BibitemOpen
  \bibfield  {author} {\bibinfo {author} {\bibfnamefont {A.}~\bibnamefont
  {Rubio}}, \bibinfo {author} {\bibfnamefont {J.~L.}\ \bibnamefont {Corkill}},
  \bibinfo {author} {\bibfnamefont {M.~L.}\ \bibnamefont {Cohen}}, \bibinfo
  {author} {\bibfnamefont {E.~L.}\ \bibnamefont {Shirley}},\ and\ \bibinfo
  {author} {\bibfnamefont {S.~G.}\ \bibnamefont {Louie}},\ }\bibfield  {title}
  {\bibinfo {title} {Quasiparticle band structure of {AlN} and {GaN}},\ }\href
  {https://doi.org/10.1103/PhysRevB.48.11810} {\bibfield  {journal} {\bibinfo
  {journal} {Phys. Rev. B}\ }\textbf {\bibinfo {volume} {48}},\ \bibinfo
  {pages} {11810--11816} (\bibinfo {year} {1993})}\BibitemShut {NoStop}%
\bibitem [{\citenamefont {Onida}\ \emph {et~al.}(2002)\citenamefont {Onida},
  \citenamefont {Reining},\ and\ \citenamefont {Rubio}}]{GWreview}%
  \BibitemOpen
  \bibfield  {author} {\bibinfo {author} {\bibfnamefont {G.}~\bibnamefont
  {Onida}}, \bibinfo {author} {\bibfnamefont {L.}~\bibnamefont {Reining}},\
  and\ \bibinfo {author} {\bibfnamefont {A.}~\bibnamefont {Rubio}},\ }\bibfield
   {title} {\bibinfo {title} {Electronic excitations: density-functional versus
  many-body {Green's}-function approaches},\ }\href
  {https://doi.org/10.1103/RevModPhys.74.601} {\bibfield  {journal} {\bibinfo
  {journal} {Rev. Mod. Phys.}\ }\textbf {\bibinfo {volume} {74}},\ \bibinfo
  {pages} {601--659} (\bibinfo {year} {2002})}\BibitemShut {NoStop}%
\bibitem [{\citenamefont {Marini}\ \emph {et~al.}(2009)\citenamefont {Marini},
  \citenamefont {Hogan}, \citenamefont {Gr{\"u}ning},\ and\ \citenamefont
  {Varsano}}]{Yambo1}%
  \BibitemOpen
  \bibfield  {author} {\bibinfo {author} {\bibfnamefont {A.}~\bibnamefont
  {Marini}}, \bibinfo {author} {\bibfnamefont {C.}~\bibnamefont {Hogan}},
  \bibinfo {author} {\bibfnamefont {M.}~\bibnamefont {Gr{\"u}ning}},\ and\
  \bibinfo {author} {\bibfnamefont {D.}~\bibnamefont {Varsano}},\ }\bibfield
  {title} {\bibinfo {title} {Yambo: an ab initio tool for excited state
  calculations},\ }\href {https://doi.org/10.1016/j.cpc.2009.02.003} {\bibfield
   {journal} {\bibinfo  {journal} {Comput. Phys. Commun.}\ }\textbf {\bibinfo
  {volume} {180}},\ \bibinfo {pages} {1392--1403} (\bibinfo {year}
  {2009})}\BibitemShut {NoStop}%
\bibitem [{\citenamefont {Sangalli}\ \emph {et~al.}(2019)\citenamefont
  {Sangalli}, \citenamefont {Ferretti}, \citenamefont {Miranda}, \citenamefont
  {Attaccalite}, \citenamefont {Marri}, \citenamefont {Cannuccia},
  \citenamefont {Melo}, \citenamefont {Marsili}, \citenamefont {Paleari},
  \citenamefont {Marrazzo}, \citenamefont {Prandini}, \citenamefont
  {Bonf{\`{a}}}, \citenamefont {Atambo}, \citenamefont {Affinito},
  \citenamefont {Palummo}, \citenamefont {Molina-S{\'{a}}nchez}, \citenamefont
  {Hogan}, \citenamefont {Grüning}, \citenamefont {Varsano},\ and\
  \citenamefont {Marini}}]{Yambo2}%
  \BibitemOpen
  \bibfield  {author} {\bibinfo {author} {\bibfnamefont {D.}~\bibnamefont
  {Sangalli}}, \bibinfo {author} {\bibfnamefont {A.}~\bibnamefont {Ferretti}},
  \bibinfo {author} {\bibfnamefont {H.}~\bibnamefont {Miranda}}, \bibinfo
  {author} {\bibfnamefont {C.}~\bibnamefont {Attaccalite}}, \bibinfo {author}
  {\bibfnamefont {I.}~\bibnamefont {Marri}}, \bibinfo {author} {\bibfnamefont
  {E.}~\bibnamefont {Cannuccia}}, \bibinfo {author} {\bibfnamefont
  {P.}~\bibnamefont {Melo}}, \bibinfo {author} {\bibfnamefont {M.}~\bibnamefont
  {Marsili}}, \bibinfo {author} {\bibfnamefont {F.}~\bibnamefont {Paleari}},
  \bibinfo {author} {\bibfnamefont {A.}~\bibnamefont {Marrazzo}}, \bibinfo
  {author} {\bibfnamefont {G.}~\bibnamefont {Prandini}}, \bibinfo {author}
  {\bibfnamefont {P.}~\bibnamefont {Bonf{\`{a}}}}, \bibinfo {author}
  {\bibfnamefont {M.~O.}\ \bibnamefont {Atambo}}, \bibinfo {author}
  {\bibfnamefont {F.}~\bibnamefont {Affinito}}, \bibinfo {author}
  {\bibfnamefont {M.}~\bibnamefont {Palummo}}, \bibinfo {author} {\bibfnamefont
  {A.}~\bibnamefont {Molina-S{\'{a}}nchez}}, \bibinfo {author} {\bibfnamefont
  {C.}~\bibnamefont {Hogan}}, \bibinfo {author} {\bibfnamefont
  {M.}~\bibnamefont {Grüning}}, \bibinfo {author} {\bibfnamefont
  {D.}~\bibnamefont {Varsano}},\ and\ \bibinfo {author} {\bibfnamefont
  {A.}~\bibnamefont {Marini}},\ }\bibfield  {title} {\bibinfo {title}
  {Many-body perturbation theory calculations using the {Yambo} code},\ }\href
  {https://doi.org/10.1088/1361-648x/ab15d0} {\bibfield  {journal} {\bibinfo
  {journal} {J. Phys.: Condens. Matter}\ }\textbf {\bibinfo {volume} {31}},\
  \bibinfo {pages} {325902} (\bibinfo {year} {2019})}\BibitemShut {NoStop}%
\bibitem [{\citenamefont {Ceperley}\ and\ \citenamefont {Alder}(1980)}]{LDA}%
  \BibitemOpen
  \bibfield  {author} {\bibinfo {author} {\bibfnamefont {D.~M.}\ \bibnamefont
  {Ceperley}}\ and\ \bibinfo {author} {\bibfnamefont {B.~J.}\ \bibnamefont
  {Alder}},\ }\bibfield  {title} {\bibinfo {title} {Ground state of the
  electron gas by a stochastic method},\ }\href
  {https://doi.org/10.1103/PhysRevLett.45.566} {\bibfield  {journal} {\bibinfo
  {journal} {Phys. Rev. Lett.}\ }\textbf {\bibinfo {volume} {45}},\ \bibinfo
  {pages} {566--569} (\bibinfo {year} {1980})}\BibitemShut {NoStop}%
\bibitem [{\citenamefont {Fuchs}\ \emph {et~al.}(2008)\citenamefont {Fuchs},
  \citenamefont {R\"odl}, \citenamefont {Schleife},\ and\ \citenamefont
  {Bechstedt}}]{FuchsBindingEnergy}%
  \BibitemOpen
  \bibfield  {author} {\bibinfo {author} {\bibfnamefont {F.}~\bibnamefont
  {Fuchs}}, \bibinfo {author} {\bibfnamefont {C.}~\bibnamefont {R\"odl}},
  \bibinfo {author} {\bibfnamefont {A.}~\bibnamefont {Schleife}},\ and\
  \bibinfo {author} {\bibfnamefont {F.}~\bibnamefont {Bechstedt}},\ }\bibfield
  {title} {\bibinfo {title} {Efficient $\mathcal{O}({N}^{2})$ approach to solve
  the {Bethe-Salpeter} equation for excitonic bound states},\ }\href
  {https://doi.org/10.1103/PhysRevB.78.085103} {\bibfield  {journal} {\bibinfo
  {journal} {Phys. Rev. B}\ }\textbf {\bibinfo {volume} {78}},\ \bibinfo
  {pages} {085103} (\bibinfo {year} {2008})}\BibitemShut {NoStop}%
\bibitem [{\citenamefont {Yoon}\ \emph {et~al.}(1996)\citenamefont {Yoon},
  \citenamefont {Wake},\ and\ \citenamefont {Wolfe}}]{Yoon1996}%
  \BibitemOpen
  \bibfield  {author} {\bibinfo {author} {\bibfnamefont {H.~W.}\ \bibnamefont
  {Yoon}}, \bibinfo {author} {\bibfnamefont {D.~R.}\ \bibnamefont {Wake}},\
  and\ \bibinfo {author} {\bibfnamefont {J.~P.}\ \bibnamefont {Wolfe}},\
  }\bibfield  {title} {\bibinfo {title} {Effect of exciton-carrier
  thermodynamics on the {GaAs} quantum well photoluminescence},\ }\href
  {https://doi.org/10.1103/PhysRevB.54.2763} {\bibfield  {journal} {\bibinfo
  {journal} {Phys. Rev. B}\ }\textbf {\bibinfo {volume} {54}},\ \bibinfo
  {pages} {2763--2774} (\bibinfo {year} {1996})}\BibitemShut {NoStop}%
\end{thebibliography}
\providecommand{\noopsort}[1]{}\providecommand{\singleletter}[1]{#1}%

\end{document}